\documentclass[journal,twoside,web]{ieeecolor}
\usepackage{generic}

\usepackage{float}
\usepackage{booktabs}
\usepackage{threeparttable}
\usepackage{siunitx}

\usepackage{tikz}
\usetikzlibrary{shapes, arrows, positioning}
\usepackage{lipsum} 

\usepackage{cite}

\usepackage{ifpdf}
\usepackage{pgfplots}
\pgfplotsset{compat=1.18} 

\usepackage{graphicx}
\ifpdf
  \graphicspath{ {./fig/} }
  \DeclareGraphicsExtensions{.pdf,.jpeg,.png}
\else
  \graphicspath{ {./fig/} }
  \DeclareGraphicsExtensions{.eps}
\fi

\usepackage{amsmath,amssymb,amsfonts}
\usepackage{algorithm}
\usepackage{algpseudocode}
\usepackage{array}
\usepackage{tabularx}
\usepackage{textcomp}
\usepackage{subcaption}

\usepackage{hyperref}
\hypersetup{
    colorlinks=true,
    linkcolor=blue,
    filecolor=magenta,      
    urlcolor=cyan,
    pdftitle={Overleaf Example},
    pdfpagemode=UseOutlines,
    }

\usepackage{url}

\def\BibTeX{{\rm B\kern-.05em{\sc i\kern-.025em b}\kern-.08em
    T\kern-.1667em\lower.7ex\hbox{E}\kern-.125emX}}
\begin{document}
\title{mmWave Radar for Sit-to-Stand Analysis: A Comparative Study with Wearables and Kinect}
\author{Shuting Hu, Peggy Ackun, Xiang Zhang, $\text{Siyang Cao}^{*}$, Jennifer Barton, Melvin G. Hector, Mindy J. Fain and Nima Toosizadeh
\thanks{$* \text{Contact author}$: S. Cao is with the Department of Electrical and Computer Engineering, The University of Arizona, Tucson, AZ, 85721 USA. e-mail: (caos@arizona.edu).}
\thanks{S. Hu is with the Department of Electrical and Computer Engineering, The University of Arizona, Tucson, AZ, 85721 USA. e-mail: (shutinghu@arizona.edu).}
\thanks{X. Zhang is with Biosystems Analytics, The Univerity of Arizona, Tucson, AZ, 85721 USA . e-mail: (xiangzhang@arizona.edu).}
\thanks{P. Ackun and J. Barton are with the Department of Biomedical Engineering, The University of Arizona, Tucson, AZ, 85721 USA. e-mail: (packun@arizona.edu, barton@arizona.edu).}
\thanks{M.G. Hector and M.J. Fain are with the Department of Medicine, The University of Arizona, Tucson, AZ, 85724 USA. e-mail: (Melvin.hector@banner.health.com, MFain@aging.arizona.edu).}
\thanks{N. Toosizadeh is with the Department of Rehabilitation and Movement Sciences, Rutgers Health, Rutgers University. e-mail: (nima.toosizadeh@rutgers.edu).}
}

\maketitle

\begin{abstract}
This study explores a novel approach for analyzing Sit-to-Stand (STS) movements using millimeter-wave (mmWave) radar technology. The goal is to develop a non-contact sensing, privacy-preserving, and all-day operational method for healthcare applications, including fall risk assessment. We used a 60GHz mmWave radar system to collect radar point cloud data, capturing STS motions from 45 participants. By employing a deep learning pose estimation model, we learned the human skeleton from Kinect built-in body tracking and applied Inverse Kinematics (IK) to calculate joint angles, segment STS motions, and extract commonly used features in fall risk assessment. Radar extracted features were then compared with those obtained from Kinect and wearable sensors. The results demonstrated the effectiveness of mmWave radar in capturing general motion patterns and large joint movements (e.g., trunk). Additionally, the study highlights the advantages and disadvantages of individual sensors and suggests the potential of integrated sensor technologies to improve the accuracy and reliability of motion analysis in clinical and biomedical research settings.

\end{abstract}

\begin{IEEEkeywords}
Sit-to-Stand, mmWave Radar, Fall Risk Assessment, Motion Analysis, Healthcare, Biomedical.
\end{IEEEkeywords}

\begin{tikzpicture}[remember picture, overlay]
\node[anchor=south, yshift=0.5cm] at (current page.south) {\footnotesize This work has been submitted to the IEEE for possible publication. Copyright may be transferred without notice, after which this version may no longer be accessible.};
\end{tikzpicture}

\section{Introduction}

\IEEEPARstart{H}{uman} motion analysis plays a pivotal role in fields such as biomechanics, rehabilitation, and sports science. Accurate assessment of human movement provides valuable insights into physical abilities, aids in diagnosing health conditions, and monitors rehabilitation progress. Various sensing technologies have been employed for motion analysis, including wearable sensors~\cite{bohannon2010sit}, vision-based systems~\cite{matthew2019estimating, morgan2023automated}, and Radio Frequency (RF)-based sensors~\cite{yan2024person, yang2022rethinking, jin2020mmfall, zhang2018real}.

Wearable sensors typically involve attaching Inertial Measurement Units (IMUs) to the body to detect limb movements. While they are popular due to their ability to capture subtle movements, they come with limitations such as discomfort during prolonged wear, the need for frequent recharging, especially among older adults who may forget to wear them due to cognitive decline~\cite{cuesta2010use}.
Vision-based systems use cameras to analyze motion by detecting changes in body pixels within images. Although they can provide detailed motion capture, they are sensitivity to lighting conditions, and have privacy concerns due to the recording of identifiable visual information. Depth sensors, like Microsoft Kinect, mitigate some privacy issues by not capturing detailed facial features but still face limitations related to line-of-sight and environmental conditions~\cite{poppe2007vision}.
RF-based systems, including those using WiFi and radar sensors, employ transmitting and receiving antennas to recognize activities by detecting signal changes caused by human movement. Early radar-based analyses leveraged micro-Doppler signatures to extract motion patterns~\cite{wang2014quantitative, seifert2019toward, saho2020evaluation}, but capturing comprehensive joint data for full-body motion analysis remains challenging.

The Sit-to-Stand (STS) transition is a fundamental movement in daily life and a vital indicator of an individual's lower limb strength, balance, and overall functional mobility. Analyzing STS movements has revealed wide-ranging applications in biomedical disease assessment, including Chronic Obstructive Pulmonary Disease (COPD), stroke, Multiple Sclerosis (MS), and osteoporosis~\cite{vaidya2017sit, bohannon20191}.

Despite the potential benefits, the application of millimeter-wave (mmWave) radar in STS motion analysis remains underexplored. Challenges persist in accurately capturing, processing, and interpreting radar data to extract meaningful motion patterns. To address this, our work builds on our previous study and employs the mmPose-FK model~\cite{hu2024mmpose}, transforming mmWave radar point cloud data into a 17-joint human skeleton. This approach enables joint-level feature extraction and signal-level comparisons with other sensors. This is the first work to comprehensively analyze STS movements using mmWave radar point cloud data. Our contributions are:

\begin{enumerate} 
\item We developed a mmWave radar-based system to capture STS movements.
\item We designed algorithms to process and extract key STS features.
\item We validated our approach compared with Kinect and wearable sensors.
\end{enumerate}

The remainder of this paper is organized as follows: Section II reviews related work in STS motion analysis and sensor technologies. Section III details the proposed methodology. Section IV presents the experiments, results, and discussion. Finally, Section V concludes the paper.

\section{Related Work}

This section reviews the literature related to STS motion analysis, including its characteristics, healthcare applications, and the sensor technologies employed.

\subsection{Sit-to-Stand Motion Analysis}


The STS movement is a fundamental activity of daily living and a key indicator of an individual's functional mobility and balance~\cite{whitney2005clinical}. Early research by Nuzik et al.\cite{nuzik1986sit} and Schenkman et al.\cite{schenkman1990whole} provided foundational analyses of STS biomechanics, identifying phases of the movement and quantifying joint angles, velocities, and torques. Van Lummel et al.~\cite{van2013automated} examined parameters such as phase durations and maximum angular velocities, revealing significant differences in movement patterns between age groups.

Analyzing STS movements has widespread applications in healthcare. STS performance has been linked to fall risk in older adults. Cheng et al.\cite{cheng1998sit} identified features like rate of rise in force and postural sway during STS that correlate with fall risk. Kera and Kinoshita\cite{kera2020association} studied the relationship between ground reaction forces during STS and fall incidence, suggesting that these measurements could predict future falls.
In stroke patients, STS analysis can provide insights into motor recovery. Mao et al.\cite{mao2018crucial} analyzed kinematic changes during STS in stroke survivors, highlighting delays and altered movement phases. Mentiplay et al.\cite{mentiplay2020five} found significant correlations between lower-limb muscle strength, balance, and five-times STS performance post-stroke.

For patients with COPD, STS tests can assess functional exercise capacity. Crook et al.\cite{crook2017multicentre} validated the one-minute STS test for COPD patients, demonstrating reliability in measuring functional capacity. Medina-Mirapeix et al.\cite{medina2022prognostic} and Spence et al.~\cite{spence2023one} further confirmed the prognostic value and feasibility of STS tests in COPD management.
STS performance is also relevant in conditions like osteoarthritis and sarcopenia. Boswell et al.\cite{boswell2023smartphone} used smartphone videos for STS analysis, linking movement parameters to osteoarthritis diagnosis. Yee et al.\cite{yee2021performance} associated STS performance with functional fitness in diagnosing sarcopenia.
In patients with MS, STS tests help assess lower extremity function. Bowser et al.\cite{bowser2015sit} examined STS mechanics in MS patients, suggesting that rehabilitation focused on leg extensor strength could improve performance. Zheng et al.\cite{zheng2023validity} validated the 30-second STS test for assessing lower limb function in MS.

Recent studies have explored STS tests in telehealth settings. Bowman et al.\cite{bowman2023feasibility} demonstrated the feasibility of the 30-second STS test via telehealth for oncology patients, enabling remote assessment of lower limb strength. Van Loon et al.\cite{van2023self} validated the five-times STS test at home for cancer survivors, providing a reliable tool for assessing frailty.

\subsection{Sensor Technologies for STS Analysis}
The Centers for Disease Control and Prevention's (CDC) Stopping Elderly Accidents, Deaths, and Injuries (STEADI) initiative~\cite{bergen2019cdc} includes the 30-second STS test, which provides a simple way to test lower body strength. However, it is limited to counting STS repetitions and may introduce human error.
Advancements in sensor technology have enabled more precise and objective measurements of STS movements:

\subsubsection{Wearable Sensors}

Wearable inertial sensors offer high-frequency, accurate data on acceleration and angular velocity. Tulipani et al.\cite{tulipani2022evaluation} used wearable sensors on the thigh to assess 30-second STS performance in patients with MS, predicting fall risk. Wang et al.\cite{wang2023high} diagnosed sarcopenia severity using STS motion data captured by wearable sensors attached to the waist and thigh, extracting numerous features to achieve high classification accuracy.

\subsubsection{Vision-Based Systems}

Vision-based systems enable detailed motion analysis. Matthew et al.\cite{matthew2019estimating} used a single depth camera to estimate dynamic effects such as joint torques and body momentum during STS, showing good concordance with full-body Motion Capture (MoCap) systems. Morgan et al.\cite{morgan2023automated} analyzed skeleton data extracted from video, estimating STS parameters like duration and speed.

However, wearable sensors can cause discomfort during prolonged wear and require frequent recharging, while vision-based systems have limitations related to line-of-sight dependency, sensitivity to lighting conditions, and potential privacy concerns.

\subsubsection{RF-Based Systems and mmWave Radar}

RF-based systems, particularly those using mmWave radar, offer non-contact sensing capabilities without privacy concerns. mmWave radar can capture movement without direct line-of-sight and is robust to lighting conditions. Saho et al. extracted kinematic information from STS movements measured with Doppler radar\cite{saho2021screening}. However, relying on micro-Doppler signatures may have limited spatial resolution~\cite{soubra2023automation}.
In recent work, Yoon et al.~\cite{yoon2024detection} detected fall risk behaviors in patients with severe mobility issues from STS movements using a Frequency Modulated Continuous Wave (FMCW) radar. They performed a range-Fast Fourier Transform (FFT) and a Doppler-FFT to acquire the range and velocity maps. Then they proposed seven features and employed the Bidirectional Long Short Term Memory (Bi-LSTM) network to detect the risk of falling. However, since these features only represent entire body information and cannot extract joint-level details, their method may be limited when comparing with other sensors at signal-level.

Advancements in Multiple-Input Multiple-Output (MIMO) technology, and Printed Circuit Board (PCB) antenna integration have enhanced the capabilities of mmWave radar and enabled more detailed motion analysis~\cite{hu2024radar}. Despite these advancements, the application of mmWave radar specifically for STS motion analysis has not been extensively studied. Our work aims to fill this gap by developing a methodology that leverages mmWave radar point cloud data for STS analysis, providing a non-contact sensing, privacy-preserving, and all-day operational method for healthcare applications.

\begin{figure}[!htbp]
    \centering
    \includegraphics[width=0.95\linewidth]{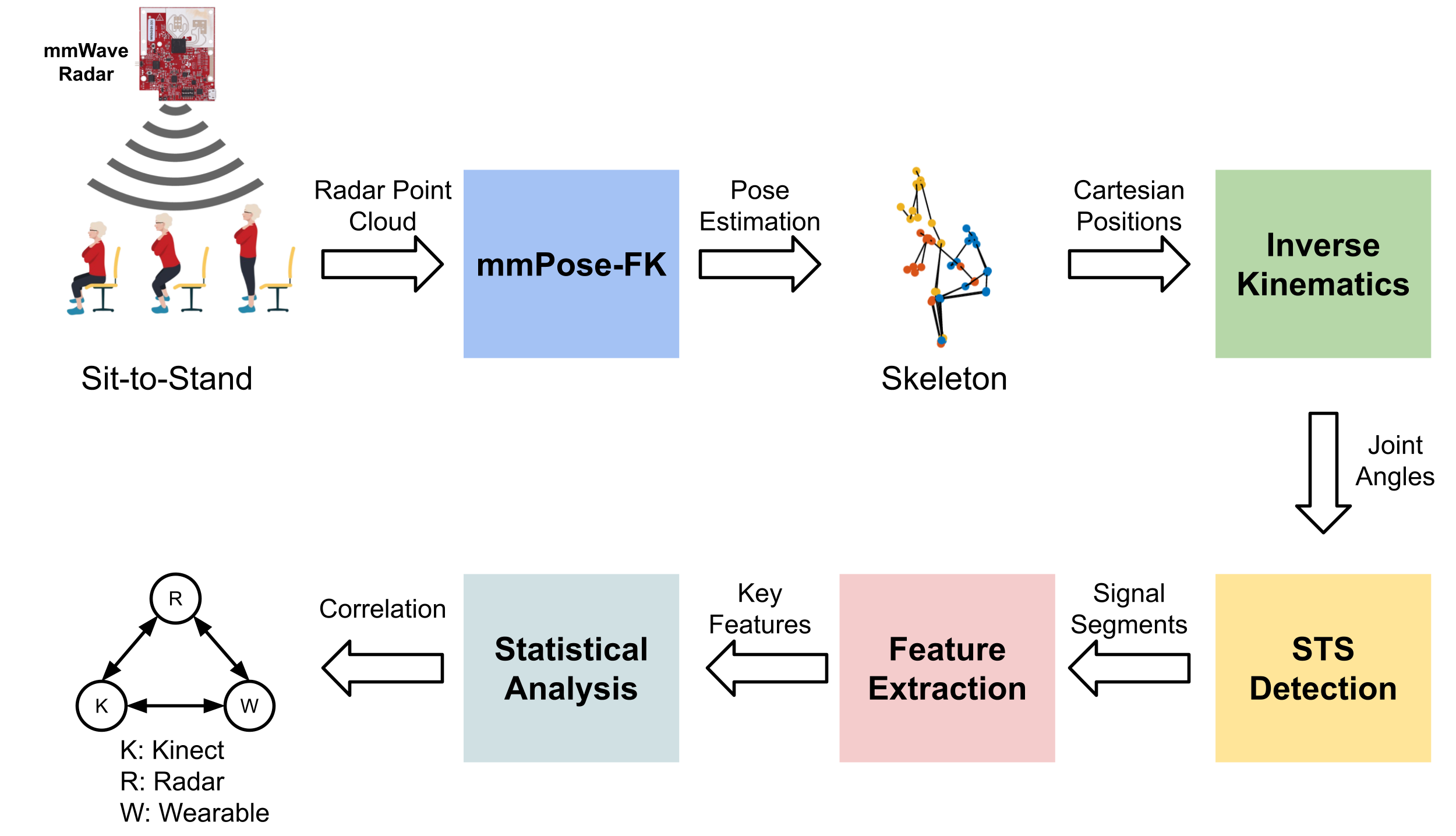}
    \caption{The proposed workflow of STS analysis using mmWave radar.}
    \label{fig:Approach}
\end{figure}

\section{Methodology}

In this section, we present our approach for conducting mmWave radar STS analysis. The workflow is shown in Fig.~\ref{fig:Approach}.

\subsection{Radar Signal Processing}


The mmWave radar operates within the 60\,GHz to 64\,GHz frequency band and is equipped with 3 transmit (TX) and 4 receive (RX) antennas. This configuration enables a 120$^\circ$ Field of View (FoV) in both azimuth and elevation directions, making the system well-suited for monitoring a standard patient room. The radar transmits FMCW signals, commonly known as chirps. Each chirp is a frequency-modulated signal where the frequency increases linearly over time. The transmitted chirp signal can be expressed as:

\begin{equation}
s_{\text{tx}}(t) = \exp\left(j 2\pi \left(f_c t + \frac{1}{2} k t^2 \right)\right),
\label{eq:tx_signal}
\end{equation}

where $f_c$ is the carrier frequency, $k$ is the chirp slope (Hz/s), and $t$ is the fast time within a chirp.
When the transmitted signal reflects off a target at range $R$, moving with radial velocity $v$, the received signal is a delayed and Doppler-shifted version of the transmitted signal:

\begin{equation}
s_{\text{rx}}(t) = \alpha \exp\left(j 2\pi \left(f_c (t - \tau) + \frac{1}{2} k (t - \tau)^2 \right)\right),
\label{eq:rx_signal}
\end{equation}

where $\alpha$ represents the attenuation factor, and $\tau = \frac{2 R}{c}$ is the time delay corresponding to the round-trip propagation time, with $c$ being the speed of light.
By mixing the received signal with a copy of the transmitted signal, we obtain the intermediate frequency (IF) signal:

\begin{equation}
\begin{aligned}
s_{\text{IF}}(t) &= s_{\text{rx}}(t) \cdot s_{\text{tx}}^*(t) \\
&= \alpha \exp\left(-j 2\pi \left( f_c \tau - k \tau t + \frac{1}{2} k \tau^2 \right) \right),
\end{aligned}
\label{eq:if_signal}
\end{equation}

where $^*$ denotes the complex conjugate. The IF signal contains frequency components corresponding to the range and velocity of the target. After sampling and digitization, we perform signal processing steps including range FFT, Doppler FFT, and angle estimation.

\subsubsection{Range Processing}

The first step is to perform FFT along the fast time (within a chirp) to obtain the range profile. The beat frequency $f_b$ corresponding to the target's range is:

\begin{equation}
f_b = k \tau = \frac{2 k R}{c}.
\label{eq:beat_frequency}
\end{equation}

By performing an $N_r$-point FFT along the fast time dimension, we can resolve targets at different ranges. The range resolution $\Delta R$ is given by:

\begin{equation}
\Delta R = \frac{c}{2 B},
\label{eq:range_resolution}
\end{equation}

where $B = k T_c$ is the bandwidth of the chirp, and $T_c$ is the chirp duration.

\subsubsection{Doppler Processing}

To estimate the target's radial velocity, we perform a Doppler FFT across multiple chirps. The phase change between consecutive chirps due to the target's motion introduces a Doppler frequency shift $f_D$:

\begin{equation}
f_D = \frac{2 v}{\lambda},
\label{eq:doppler_frequency}
\end{equation}

where $v$ is the radial velocity, and $\lambda = \frac{c}{f_c}$ is the wavelength.

By performing an $N_d$-point FFT along the slow time dimension (across chirps), we can estimate the Doppler frequency and thus the target's velocity. The velocity resolution $\Delta v$ is:

\begin{equation}
\Delta v = \frac{\lambda}{2 N_d T_c},
\label{eq:velocity_resolution}
\end{equation}

where $N_d$ is the number of chirps in a frame.

\subsubsection{Angle Estimation}

The radar employs a MIMO antenna array configuration to estimate the Angle of Arrival (AoA) of the reflected signals. By processing the signals received at different RX antennas, we can estimate the azimuth and elevation angles using techniques such as beamforming or the MUSIC algorithm.

Assuming a Uniform Linear Array (ULA) with element spacing $d$, the phase difference between adjacent antennas due to a target at angle $\theta$ is:

\begin{equation}
\Delta \phi = \frac{2\pi d \sin \theta}{\lambda}.
\label{eq:phase_difference}
\end{equation}

By analyzing the phase differences across the antenna array, we can estimate the AoA. The angular resolution $\Delta \theta$ depends on the array aperture and is given by:

\begin{equation}
\Delta \theta = \frac{\lambda}{2 L \cos \theta},
\label{eq:angular_resolution}
\end{equation}

where $L$ is the total length of the antenna array.

After calculating the range, azimuth angle, elevation angle, and velocity of the targets, we will get the radar point clouds data. In practice, the radar signal processing chain may also involve static clutter removal, and Constant False Alarm Rate (CFAR) detection algorithms.


\subsection{Pose Estimation}


\begin{figure}[!htbp]
    \centering
    \includegraphics[width=0.95\linewidth]{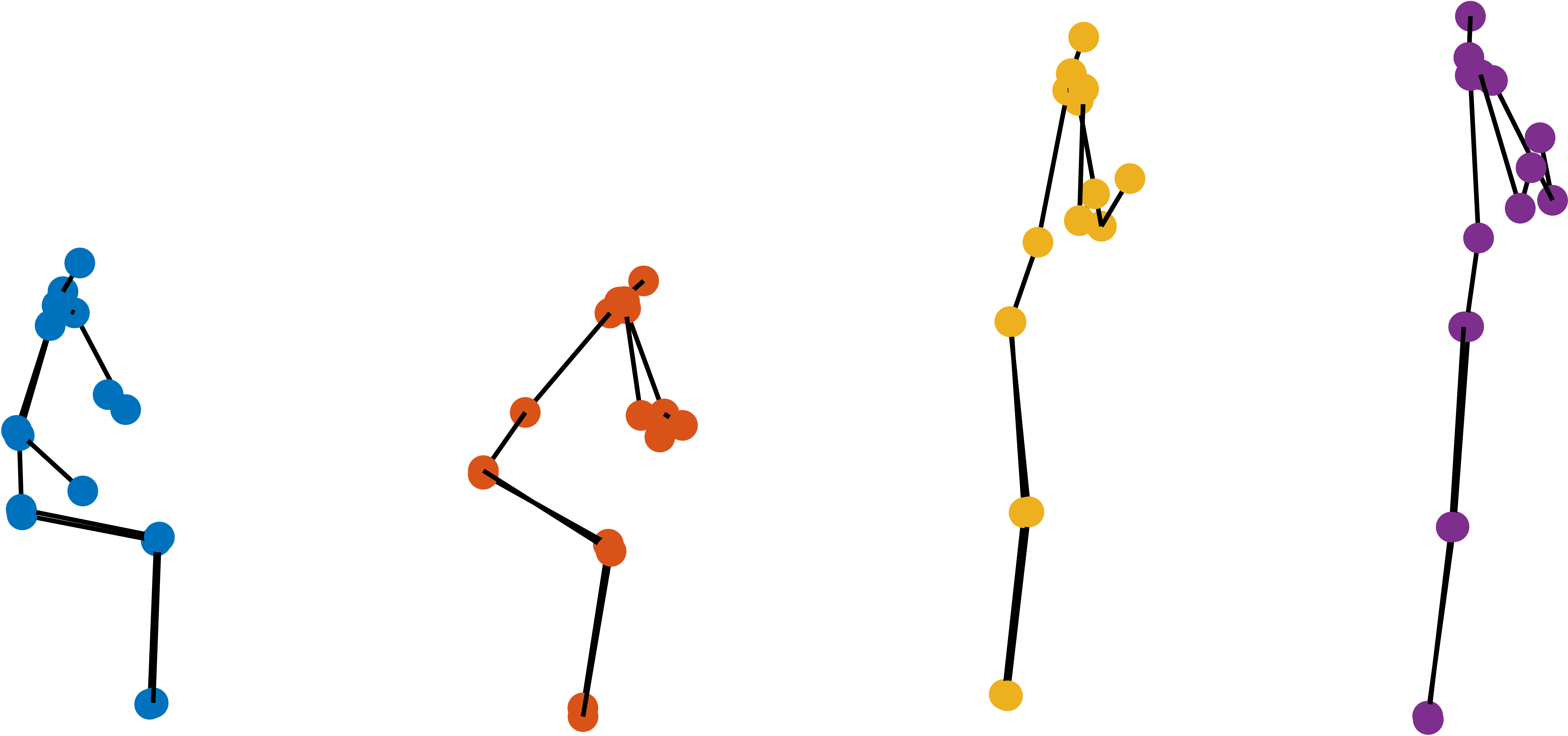}
    \caption{ Recovered STS skeleton with four phases (Sitting: blue, Flexion: orange, Extension: yellow, Standing: purple).}
    \label{fig:pose}
\end{figure}

For human motion analysis, we may choose to use radar's micro-Doppler signature, but the task of tracking limbs through micro-Doppler data poses accuracy challenges~\cite{gurbuz2024overview}. Our previous work~\cite{sengupta2020mm,sengupta2022mmpose} accurately recovered the human skeleton using radar point cloud. To further stabilize the skeleton, we used a filtering approach~\cite{hu2022stabilizing} and developed a Forward Kinematics (FK) method by constraining the skeleton within a predefined T-pose human model. Fig.~\ref{fig:pose} shows the STS skeleton predicted using the mmPose-FK model~\cite{hu2024mmpose}, where the person transitions from a sitting state, through trunk flexion and trunk extension, to a standing state.

In our mmPose-FK model, we implemented the Leave-One-Out Cross-Validation (LOOCV) strategy. Each validation iteration used data from one participant as the test set, with the remaining data forming the training set. The training data were randomly shuffled and split into an 8:2 training-validation ratio. The test data of the left-out participant was not shuffled. 


\begin{figure*}[!htbp]
    \centering
    \includegraphics[width=0.95\linewidth]{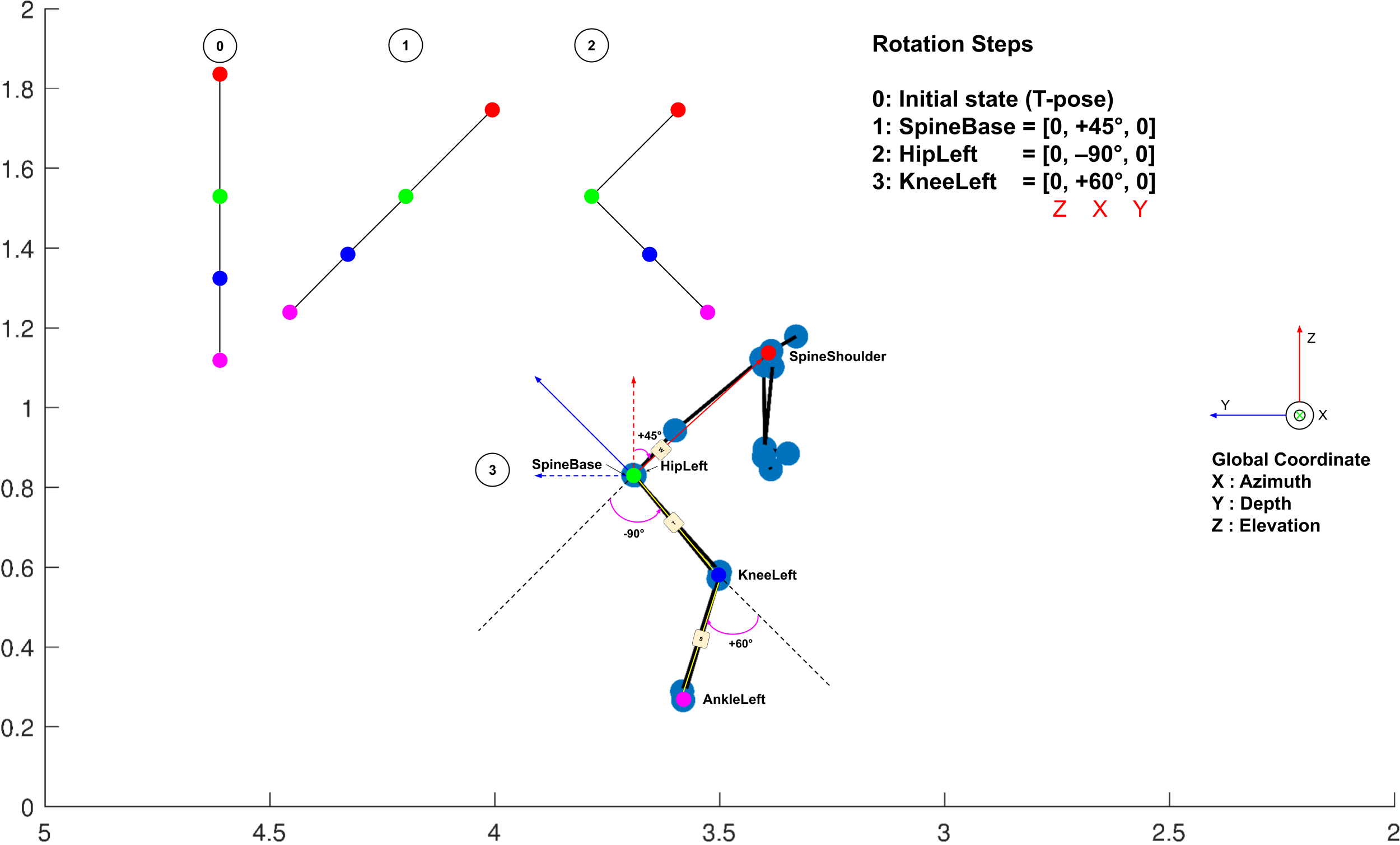}
    \caption{Joint angle calculation steps 1 to 3: All joints are positioned relative to the \textit{SpineBase}. Joint angles are relative values with respect to their parent joints. Five wearable sensors were placed on the waist (W), left and right thigh (T), left and right shank (S).}
    \label{fig:angle}
\end{figure*}

\subsection{Joint Angle Calculation}


To analyze STS motion, we apply IK to calculate the optimal joint angle rotations for each continuous skeleton frame. 
Each joint is associated with its parent in the skeletal hierarchy, creating a kinematic chain. 
All joints are positioned relative to the root joint, \textit{SpineBase}. We calculate the average bone lengths across all frames, defined as the Euclidean distances between connected joints. In the T-pose, each joint is given an initial offset vector with respect to its parent joint. 
 For example, the \textit{HipLeft} joint is defined by the vector $\mathbf{u}_{\text{HipLeft}} = [1, 0, 0]^\top$, and the \textit{KneeLeft} joint by $\mathbf{u}_{\text{KneeLeft}} = [0, -1, 0]^\top$. This offset vector is multiplied by the bone length $L$ to obtain the initial vector $\mathbf{U}_{i} \in \mathbb{R}^3$, where $i$ indexes the joint.

 
We start from the root joint (\textit{SpineBase}) and calculate its rotation matrix $\mathbf{R}_{\text{SpineBase}} \in \mathbb{R}^{3 \times 3}$, defined by the vectors of \textit{SpineBase}, \textit{HipLeft}, and \textit{SpineShoulder}. 
Euler angles $(\theta_z, \theta_x, \theta_y)$ are derived from this matrix, representing rotational components around the Z, X, and Y axes, respectively. The coordinate system used in this study is illustrated in Fig.~\ref{fig:angle}. The X-axis represents the azimuth direction and aligns with the sagittal (longitudinal) plane. The Y-axis represents the depth direction and aligns with the coronal (frontal) plane. The Z-axis represents the elevation direction and aligns with the transverse (axial) plane.
 
For the subsequent joint angle calculations, each joint's rotation is adjusted relative to its parent joint. This involves applying the inverse of the parent joint's rotation matrix $\mathbf{R}_{j} \in \mathbb{R}^{3 \times 3}$ to align the child joint $i$ to the orientation of parent joint $j$:


\begin{equation}
\mathbf{V}_{i} = \mathbf{R}_{j}^{-1} \left( \mathbf{P}_{i} - \mathbf{P}_{j} \right),
\end{equation}

where $\mathbf{P}_{i} \in \mathbb{R}^3$ is the position vector of the child joint $i$, $\mathbf{P}_{j} \in \mathbb{R}^3$ is the position vector of the parent joint $j$, $\mathbf{V}_{i} \in \mathbb{R}^3$ represents the direction vector from parent joint $j$ to child joint $i$ in the local coordinate frame.


Let $\mathbf{U}_{i} \in \mathbb{R}^3$ and $\mathbf{V}_{i} \in \mathbb{R}^3$ be the initial and current vectors of joint $i$, respectively. We normalize these vectors to obtain unit vectors:

\begin{equation}
\mathbf{u} = \frac{\mathbf{U}_{i}}{\|\mathbf{U}_{i}\|}, \quad \mathbf{v} = \frac{\mathbf{V}_{i}}{\|\mathbf{V}_{i}\|}.
\end{equation}

Compute the axis of rotation $\mathbf{w} \in \mathbb{R}^3$ and the scalar $c$ using:

\begin{equation}
\mathbf{w} = \mathbf{u} \times \mathbf{v}, \quad c = \mathbf{u} \cdot \mathbf{v},
\end{equation}

where $\mathbf{w}$ is the cross product of $\mathbf{u}$ and $\mathbf{v}$, $c$ is the dot product of $\mathbf{u}$ and $\mathbf{v}$.
The skew-symmetric matrix $\mathbf{W} \in \mathbb{R}^{3 \times 3}$ of $\mathbf{w}$ is:

\begin{equation}
    \mathbf{W} = \begin{bmatrix}
        0 & -w_3 & w_2 \\
        w_3 & 0 & -w_1 \\
        -w_2 & w_1 & 0 
    \end{bmatrix}, \\
\end{equation}


The rotation matrix $\mathbf{R}_{i} \in \mathbb{R}^{3 \times 3}$ can be computed using Rodrigues' rotation formula:

\begin{equation}
\mathbf{R}_{i} = \mathbf{I} + \mathbf{W} + \mathbf{W}^2 \frac{1 - c}{\|\mathbf{w}\|^2},
\end{equation}

where $\mathbf{I} \in \mathbb{R}^{3 \times 3}$ is the identity matrix, $\|\mathbf{w}\|$ is the magnitude of $\mathbf{w}$.


This process is repeated for each hierarchy branch and joint to construct the entire skeleton. Fig.~\ref{fig:angle} shows an example of rotations with steps: (1)~\textit{SpineBase}, (2)~\textit{HipLeft}, and (3)~\textit{KneeLeft}, transitioning from the initial T-pose to the current posture. 
We perform these operations for all $N$ frames, resulting in a time series of rotations with a shape of $[N, 12, 3]$, where $N$ is the number of frames, $12$ is the number of joints (excluding end leaf joints), $3$ represents the rotation angles around the Z, X, and Y axes.

Five wearable sensors were placed on the waist (W), left and right thigh (T), and left and right shank (S). Three sensors were aligned in the same coordinate.

\subsection{Sensors Synchronization}


In the mmPose-FK pose estimation model, the Kinect skeleton serves as the learning target for the radar. The radar skeleton represents the model's predicted output, while the Kinect skeleton serves as the corresponding ground truth, with both consisting of 17 joints. Therefore, the methods used to calculate joint angles for both the Kinect and radar skeletons are identical. However, the signals from them differ from those of the wearable sensors. For example, the wearable sensors output gyroscope and accelerometer signals for the waist, thigh, and shank body segments. We only use the gyroscope signals over time. The gyroscope outputs angular velocity and we applied similar filtering as our previous papers\cite{toosizadeh2020effect}, then derived angle rotations through integration. 

We are interested in the angles in the sagittal plane for the waist and knee joints. The radar/Kinect \textit{SpineBase} joint corresponds to the wearable's waist sensor. But for the knee joint, it requires using the angle between the thigh and shank to get the corresponding angle. Additionally, we averaged the signals from the left and right legs. The final signals we get are the trunk (waist) angle and the knee angle both in the sagittal plane. We also derived angular velocity from the angles by taking the derivative with respect to the time interval between frames (the sampling rate is 20Hz for radar/Kinect and 100Hz for wearables).



To reduce the noise in the rotation angle signal, we first applied High-Pass (HP) and Low-Pass (LP) filtering. Then, we further smoothed the signal using the Whittaker-Eilers filters described in \cite{hu2022stabilizing} for both angle and velocity. Additionally, we synchronized the signals over time. The timestamps for the radar and Kinect were recorded during the data collection, and matched during pose estimation data preprocessing stage, so they shared the same timestamps. However, the wearable device only displays the elapsed time starting from zero. To reconstruct its timestamps, our solution was based on the modified time of the stored file (assumed to be the time the last frame was written) and the frame length and sampling frequency (Fs). This method provided a reasonable approximation, but potential discrepancies between the file write time and actual time still exist. To optimize the analysis, we extracted time vectors for each sensor and determined the common time range. We then interpolated the signals to a common time base and computed the cross-correlation for the interpolated knee angle signals. By identifying the index of the maximum correlation, we determined the time shift (lag), subsequently adjusting the wearable time vector accordingly.


\subsection{STS Segmentation}

\begin{figure}[!htbp]
    \centering
    \subfloat[]{\includegraphics[width=0.95\linewidth]{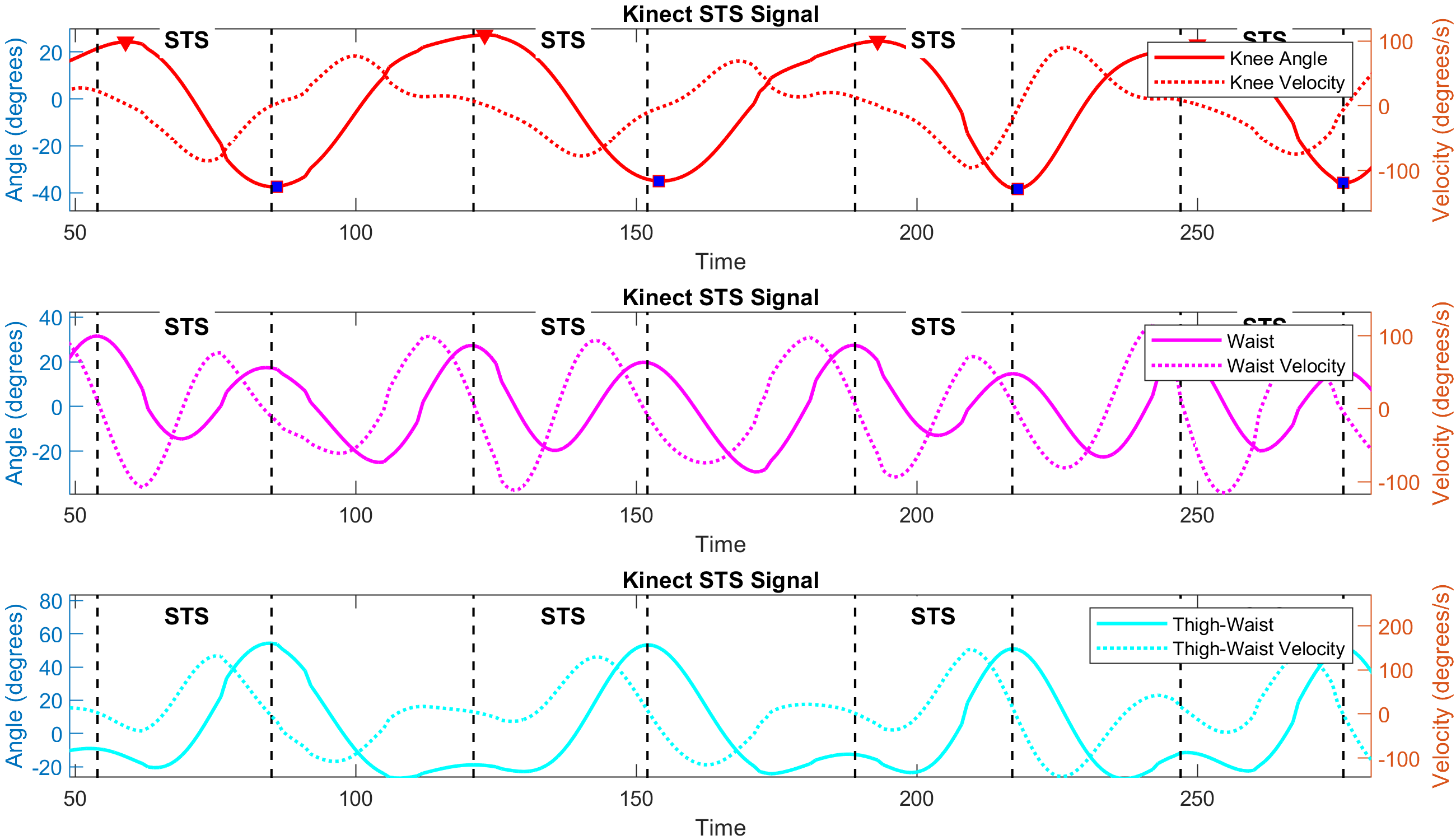}}\\
    \subfloat[]{\includegraphics[width=0.95\linewidth]{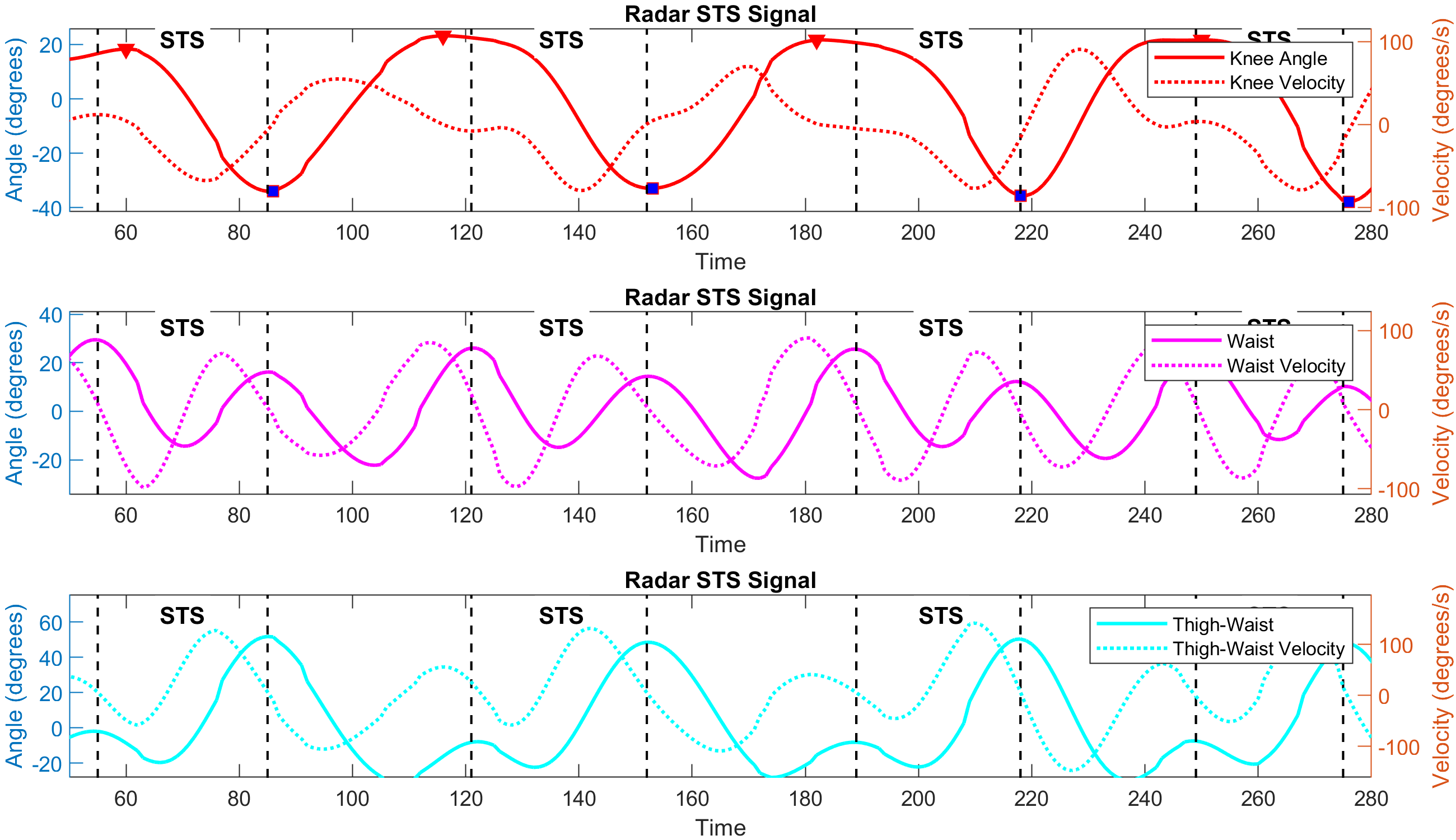}}\\
    \subfloat[]{\includegraphics[width=0.95\linewidth]{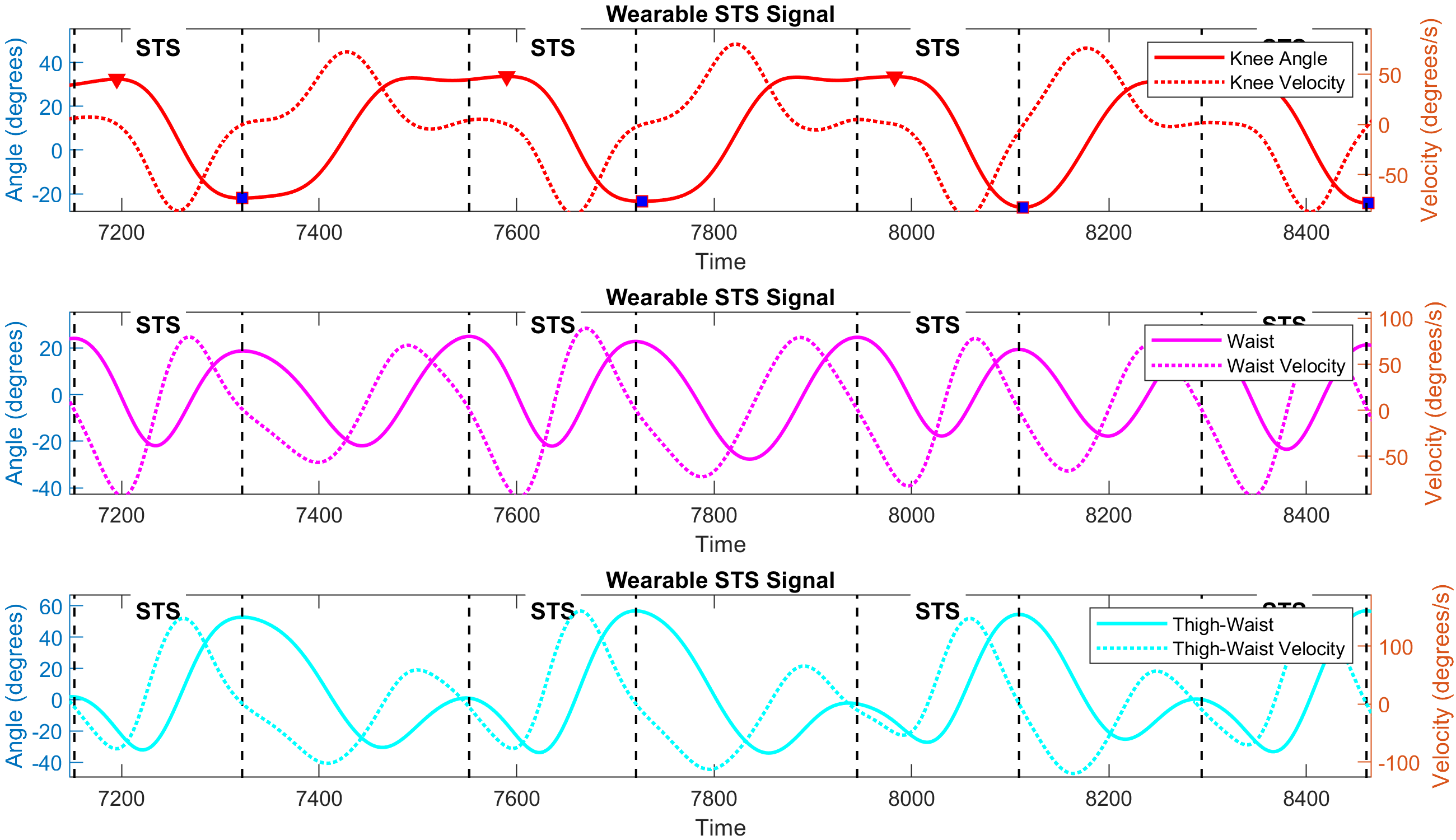}}
    \caption{Based on the synchronized signals, segmentation is performed for (a) Kinect, (b) Radar, and (c) Wearable.}
    \label{fig:SegImg}
\end{figure}

Subsequently, we developed an automated algorithm for STS repetition identification for Kinect, radar, and wearable sensors. As shown in Fig.~\ref{fig:SegImg}, high peaks (sitting - red triangle) and low peaks (standing - blue square) were detected based on the knee angle to identify the initial transitions. We iterated over each high peak to find the corresponding low peak. The left boundary of the STS segment was updated based on the high peak positions of the waist angle signal, and the right boundary was updated based on the high peak positions of the thigh-waist angle signal. Each STS repetition is indicated between two vertical dashed lines. The segmentation results show a high degree of similarity across the three sensors, which is a signal-level comparison. When handling data from different participants, we encountered various situations, such as knee joint jitter, the wearing direction of a particular wearable sensor being reversed, and so on. We put a lot of effort into developing algorithms and solving these problems.

\subsection{Feature Extraction}

From the STS segments, we calculated the duration between the STS transitions. \textit{KneeROM} is equal to the maximum knee angle minus the minimum knee angle within this repetition. 
Similarly, \textit{TrunkROM} is the maximum trunk angle minus the minimum trunk angle. The STS trunk angle changes has two trends, it first decreases and then increases during one STS cycle. 
The trunk velocity shows two peaks, the first peak corresponds to trunk flexion, referred to as \textit{TrunkFlexionPeakVelocity}, and the second peak corresponds to trunk extension, referred to as \textit{TrunkExtensionPeakVelocity}. 

During signal segmentation, we also calculated the angle between the waist and thigh to more accurately identify the end (standing) point of STS, and calculated \textit{WaistThighROM}.
The signal segmentation and feature extraction for Kinect, radar, and wearable sensors were performed separately. The number of repetitions detected by each sensor may not be exactly the same, as noise in one sensor's signal may prevent detection of all repetitions. Therefore, we also saved the start timestamp of each repetition and match the extracted features using their start timestamps of three sensors.
First, we read the saved features files (.xlsx) from the three sensors for each participant, using the radar data as the reference. We then found the closest matching start time in the Kinect data, and likewise in the wearable data, if the time difference is within an acceptable range (0.5 seconds). If such matches were found, we save these matched features. This process ensures that the number of repetitions retained for each participant and each sensor is the same for subsequent data analysis. We then organized the features of all participants into a long table format for data analysis.

\subsection{Statistical Analysis}
In this study, we aim to explore the possibility of mmWave radar for STS analysis and evaluate its performance by comparing with Kinect and wearable sensors. 
 We noticed that there were some abnormal data points (outliers) in the extracted features. To ensure the validity of our statistical comparisons, we applied the Z-score method to assess data normality and identify outliers. This method calculates the Z-score for each data point, indicating how many standard deviations the point is from the mean. Any data points with a Z-score greater than 3 or less than -3 were considered outliers and were subsequently removed from the analysis. To assess the agreement among the extracted features, we used the Intraclass Correlation Coefficient (ICC) method and the Bland-Altman test, as recommended in~\cite{toosizadeh2020effect}.

The ICC values were calculated to assess the reliability and agreement between three different measurement methods (Kinect, radar, and wearable) for various feature measurements. Specifically, we used a two-way mixed-effects model of ICC because the subjects in this study are considered random effects, and each subject was measured by the same set of sensors. ICC values were calculated pairwise between radar and wearable sensors (R-W), Kinect and radar sensors (K-R), and Kinect and wearable sensors (K-W). A higher ICC value (close to 1) indicates a strong agreement between the measures from the pair of sensors, suggesting that the measurements are highly correlated. Conversely, a lower ICC value (close to 0) indicates a weak agreement between the measures from the pair of sensors. 
 We also analyzed the mean difference between measures from two different sensors and the 95\% confidence interval following the Bland and Altman test.

\section{Experiments and Results}

\begin{table}[htbp] \small
\centering
\caption{Demographic summary}
\label{tab:demographic_summary}
\begin{tabular}{@{}lcccc@{}}
\toprule
Metric & Mean & Median & Min & Max \\ \midrule
Age & 27.04 & 21.00 & 18.00 & 79.00 \\
Height (ft) & 5.61 & 5.58 & 5.00 & 6.42 \\
Weight (lb) & 163.53 & 160.00 & 119.00 & 290.00 \\ \addlinespace
Gender & \multicolumn{4}{l}{Female: 20} \\
 & \multicolumn{4}{l}{Male: 25} \\ \addlinespace
Racial & \multicolumn{4}{l}{American Indian/Alaska Native: 0} \\
 & \multicolumn{4}{l}{Asian: 10} \\
 & \multicolumn{4}{l}{Native Hawaiian or Other Pacific Islander: 1} \\
 & \multicolumn{4}{l}{Black or African American: 2} \\
 & \multicolumn{4}{l}{White: 24} \\
 & \multicolumn{4}{l}{More than One Race: 5} \\
 & \multicolumn{4}{l}{Unknown or Not Reported: 3} \\ \addlinespace
Ethnic & \multicolumn{4}{l}{Not Hispanic or Latino: 29} \\
 & \multicolumn{4}{l}{Hispanic or Latino: 15} \\
 & \multicolumn{4}{l}{Unknown/Not Reported Ethnicity: 1} \\
\bottomrule
\end{tabular}
\end{table}

\subsection{Datasets}

Our experiments used the IWR6843ISK-ODS mmWave radar from Texas Instruments (TI) and the Microsoft Azure Kinect. We employed the built-in body tracking feature from Kinect to obtain the joint's position. We designed a 3D-printed bracket to hold the radar and Kinect together. To ensure a precise match between the radar point clouds and Kinect joint positions, we associated their coordinate systems through calibration using rotation and transformation matrices.
Five wearable motion sensors (LEGSys, BioSensics, Boston, MA, USA) with tri-axial gyroscopes were affixed to each shank on both legs, the thighs on both legs and the waist to estimate angular velocity. 
A total of 45 healthy adults (18 or older) were recruited for this study. 
 Table~\ref{tab:demographic_summary} provides a summary of the demographic information for all subjects, showing the generalizability of our dataset.



Upon arrival, participants signed a consent form and completed the Montreal Cognitive Assessment (MoCA) questionnaire to assess their cognitive ability. The MoCA has a maximum score of 30, and participants were required to score $\geq$ 21 to qualify.
Participants aged 65 or older were additionally required to complete the Centers for Medicare and Medicaid Services Hierarchical Condition Category (CMS-HCC) questionnaire. This assessment evaluates comorbidities by identifying chronic conditions that could affect participants' overall health and potentially influence study outcomes. While future fall risk assessments should account for the impact of comorbidities, the present study focuses on comparing sensor data. The presence of chronic conditions among some participants is acknowledged but not included in the analysis.

A straight-backed chair without armrests (17” seat height) was positioned 3.5 meters from the radar and Kinect sensors. Participants were given the following instructions:

\begin{enumerate} 
\item Sit in the center of the chair.
\item Keep your feet flat on the floor. 
\item Cross your arms on the opposite shoulders. 
\item Maintain a straight back. 
\item On `Go," rise to a full standing position, then sit back down again. 
\end{enumerate}

Data collection was carried out in various environments at The University of Arizona, including the Electrical and Computer Engineering (ECE) Lab and the Health Sciences Sensor Lab.
The procedure complied with the guidelines set by our institution's Institutional Review Board (IRB).
 This project has been reviewed and approved by the IRB or designee. The University of Arizona maintains a Federalwide Assurance (FWA) with the Office for Human Research Protections (OHRP) (FWA \#00004218). This Institution assures that all of its activities related to human subjects research, regardless of the source of support, will be guided by the Belmont Report and applicable regulations according to 45 CFR 46.111 and/or 21 CFR Part 50.

\begin{table*}[htbp] \normalsize
    \centering
    \caption{Intraclass Correlation Coefficients (ICCs) for STS features across different sensors}
    \label{tab:features}
    \begin{threeparttable}
    \begin{tabular}{llllll}
        \toprule
        Movement & Feature & ICC (K-R) & ICC (K-W) & ICC (R-W) &  \\
        \midrule
        STS & \textbf{Duration (s)} & \textbf{0.9556} & \textbf{0.9634} & \textbf{0.9526} & \\
            
            & \textbf{TrunkROM (°)} & \textbf{0.9295} & \textbf{0.7865} & \textbf{0.7218} & \\
            & \textbf{TrunkFlexionPeakVelocity (°/s)} & \textbf{0.8723} &\textbf{0.7348} & \textbf{0.6209} & \\
            & TrunkExtensionPeakVelocity (°/s) & \textbf{0.7862} & 0.5139 & 0.3787 & \\
            & WaistThighROM (°) & \textbf{0.6357} & 0.1891 & 0.1708 & \\
            & KneeROM (°) & 0.3064 & 0.0697 & 0.0456 & \\
        \bottomrule
    \end{tabular}
    \end{threeparttable}
\end{table*}

\subsection{Results of ICCs}

Following the previous steps, we extracted STS features for all participants. Table~\ref{tab:features} presents the ICC results comparing the three sensors. We focus on these specific features because previous studies have identified them as the most important for assessing fall risk during STS transitions. Additionally, the ICC between the radar and Kinect also reflects how effectively the radar learns from the Kinect, given that the Kinect's skeleton data was used as ground truth during the radar's pose estimation process.

\subsubsection{Duration}

\textit{Duration} measures the total time taken to complete the STS movement.
For this feature, the ICC values are high across all comparisons (K-R: 0.9556, K-W: 0.9634, R-W: 0.9526), indicating excellent agreement. This suggests that all three methods are consistently measuring the duration of the STS movement accurately. In other words, they all perform well in counting the number of STS movements.

\subsubsection{Trunk}
\textit{TrunkROM} measures the extent of movement of the trunk during the STS movement. The ICC values (K-R: 0.9295, K-W: 0.7865, R-W: 0.7218) indicate good to excellent agreement, suggesting that the measurement of trunk movement is relatively consistent across these sensors. 
\textit{TrunkFlexionPeakVelocity} measures the peak velocities at which the trunk moves forward during the STS movement. The ICC values indicate good agreement between all sensors (K-R: 0.8723, K-W: 0.7348, R-W: 0.6209).
\textit{TrunkExtensionPeakVelocity} measures the highest speed at which the trunk moves backward to its neutral position during the STS movement. The ICC values demonstrate good agreement between Kinect and radar, suggesting effective pose estimation for this feature. In contrast, moderate agreement with wearable sensors indicates variability in measurements. While wearable sensors are highly sensitive to small displacements and velocity changes, Kinect and radar may lose some information, potentially affecting their ability to capture subtle movements accurately.

\subsubsection{Waist-Thigh}
\textit{WaistThighROM} measures the extent of movement between the waist and thigh during the STS movement. Previous studies have suggested that the degree of trunk forward angle during STS is related to fall risk, showing different distributions between older adults and younger groups. For example, older adults may need a larger trunk angle before standing up due to insufficient strength.
The ICC values indicate moderate agreement between Kinect and radar but poor agreement with wearable sensors. While participants were instructed to keep their feet flat on the floor and their backs straight, variations in adherence to this posture can result in initial angle biases. To address this in future studies, it is important to ensure participants adopt a standardized sitting posture during wearable sensor synchronization. 

\subsubsection{Knee}
\textit{KneeROM} measures the extent of movement at the knee joint during the STS movement.
The ICC values are relatively low (K-R: 0.3064, K-W: 0.0697, R-W: 0.0456). This indicates that the measurement of knee joint motion varies significantly. The radar also fails to learn this feature well from the Kinect. Indeed, during data collection, we observed the Kinect skeleton's abnormal jittering in the lower leg. We argue that this is a significant factor contributing to the low ICC for the knee joint.

Overall, the analysis highlights that while some features (e.g., \textit{Duration}) show excellent agreement across different measurement methods, others exhibit significant variability (e.g., \textit{KneeROM}). This highlights the importance of considering the specific feature when interpreting the results for frailty and fall risk assessments.

\subsection{Results of Bland-Altman Plots}

\begin{figure*}[!htbp]
    \centering
    \includegraphics[width=0.95\textwidth]{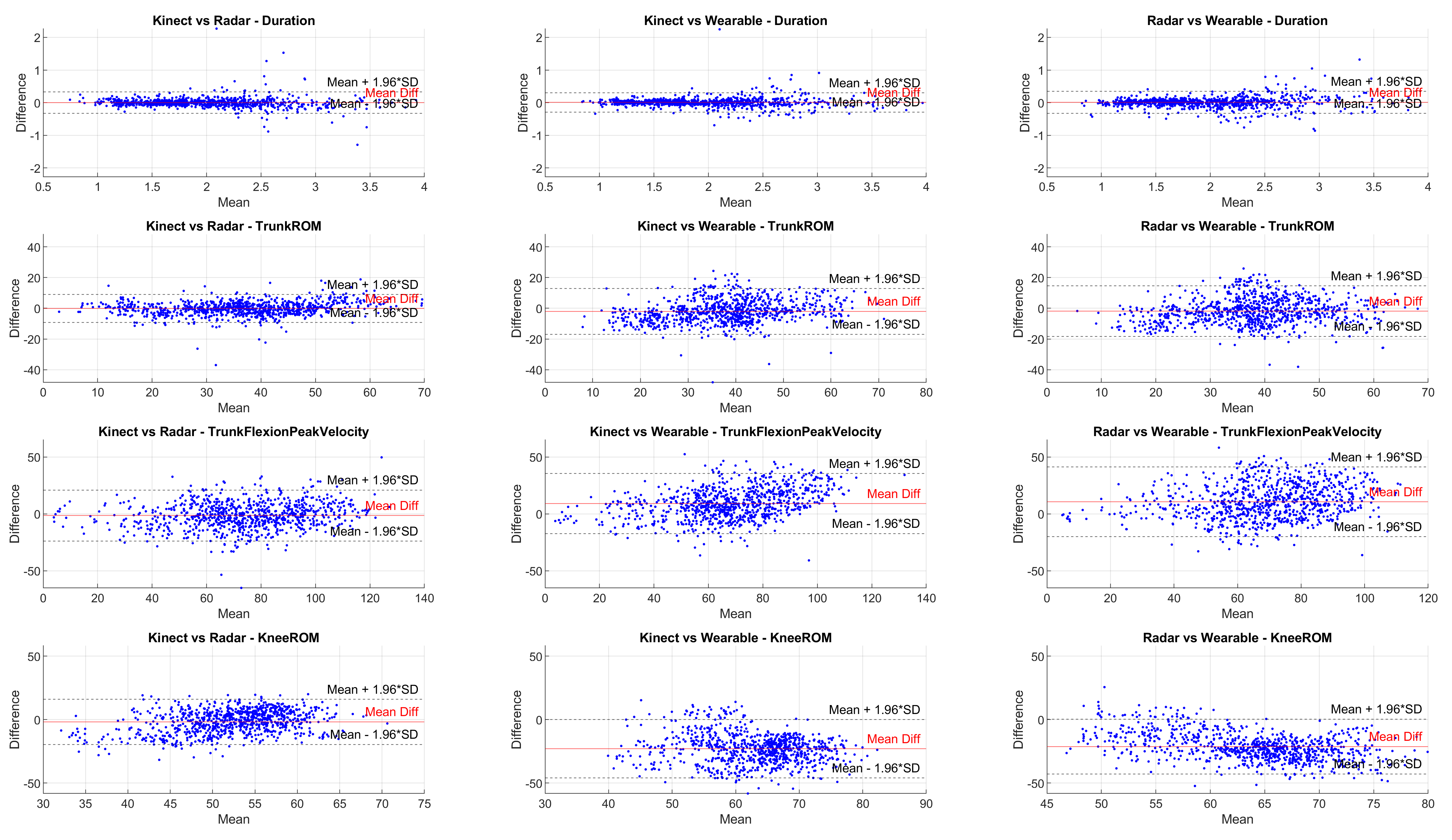}
    \caption{Bland-Altman plots showing the agreement between different measurement methods.}
    \label{fig:Statistical}
\end{figure*}

The plots illustrate the agreement between sensors for four features, as shown in Fig.~\ref{fig:Statistical}. 
 The x-axis represents the mean of the measurements, while the y-axis shows the differences between them. Each plot includes the mean difference (solid red line) and the limits of agreement (dashed black lines), defined as the mean difference ± 1.96 times the standard deviation of the differences. The Bland-Altman plots provide a clear visual summary of sensor agreement across the four features, highlighting both consistencies and discrepancies.

The top row of plots shows high agreement across all sensor pairs, with mean differences close to zero and narrow limits of agreement. The consistent clustering of points within these limits indicates minimal variability and reliable duration measurements across sensors.
The second row reveals some variability in \textit{TrunkROM} agreement. The K-R comparison demonstrates a narrow range of differences, although a few points fall outside the limits. In contrast, K-W and R-W show larger spreads and noticeable mean differences.
In the third row, K-R exhibits better agreement, while K-W and R-W comparisons show more variability, including noticeable outliers and mean differences. This aligns with the nature of velocity as a derivative of trunk angle, where even minor differences in angle measurements can amplify discrepancies.
The bottom row shows the largest variability and outliers among all features, particularly in the K-R and R-W comparisons. The greater spread and larger mean differences suggest significant inconsistencies in \textit{KneeROM} measurements. 
These findings align with the ICC results.

\subsection{Discussion}

In this paper, we focus on the STS analysis since it is a relatively simpler movement. Complex movements like Timed Up and Go (TUG) involve more dynamic and intricate motions, which present greater challenges for mmWave technology. This study serves as our initial task to validate and refine our methodology before applying it to more complex movements.

The results demonstrate that all three sensors can consistently measure the duration of STS movements, reliably counting the number of STS movements a person can complete within a fixed amount of time. When comparing the movements of specific joints, such as the trunk and knee, we observed varying levels of consistency between the sensors. For the trunk, all three sensors showed reasonable consistency in tracking trunk movements, including range of motion and peak velocity. One factor contributing to the low consistency in knee joint measurements is the significant jitter in the lower leg of the skeleton output by the Kinect, which poses difficulties for the radar learning from the Kinect.  To enable the radar system to better learn and detect joint movements, such as those of the knee, we may need to use a more accurate ground truth provider for skeleton data in future work, such as the VICON MoCap system. Besides, it is important to note that wearables were placed on the trunk, thigh, and shank, while the radar and Kinect tracked the waist, hip, and knee joints. At the start of data collection, wearable sensors require initialization (or synchronization). Inconsistent sitting postures among participants could lead to initial angle biases. Despite this, all three sensors are capable of determining the start and end of STS movements. These findings suggest that while the mmWave radar used in this study may struggle to accurately capture smaller segment due to its lower angular resolution, it still demonstrates high consistency with the other sensors in measuring trunk movements. Future studies should consider employing a more precise ground truth system to further validate these findings. In summary, while the radar may have limitations in recovering detailed movements of smaller segment, it remains a reliable tool for assessing trunk movements, showing high agreement with the Kinect and wearable sensors, particularly for trunk-related features.

Another insight we gained is the potential application of multi-sensor fusion. Each sensor has its own advantages and disadvantages in measuring movements. 
mmWave radar ensures privacy, and remains unaffected by lighting conditions, making it an excellent choice for healthcare applications.
In the meantime, we also acknowledge the limitation of radar sensors, for example, metallic objects (e.g., walking aids, wheelchairs) can significantly affect the signal, and changes in room furniture layout can alter signal reflections. Besides, moving objects, such as pets, are another common source of noise. Additionally, the mmWave radar used in the experiments has limitations in resolution. Specifically, the current azimuth and elevation angle resolution is 29 degrees. We anticipate that hardware improvements in the future will increase the angular resolution, leading to higher-quality radar point clouds of human indoor movements.
Kinect excels at recording detailed videos, providing rich visual data that can be used to analyze complex movements for post-event manual verification. Its major drawbacks are its dependency on a clear line of sight, potential for occlusions, and privacy concerns. Wearable sensors, on the other hand, are highly sensitive to small vibrations and movements, making them ideal for detecting subtle changes and fine motor movements. However, they are limited by their placement on the body and can be inconvenient for the wearer. For a comprehensive solution to more complex movements, an integrated system combining all three types of sensors would be advantageous. This hybrid approach can leverage the strengths of each sensor: radar can provide overall movement patterns, Kinect can offer detailed visual analysis, and wearables can capture fine motor details. Such a system would ensure a more robust and accurate assessment of complex movements.

In addition, we would like to note that all 45 participants in our study are healthy individuals. While we acknowledge that this may limit immediate associations with specific health conditions, recruiting healthy participants is often easier than recruiting older adults with mobility deficits or a high risk of falling. We are currently in the process of collecting data from older adults, and once we have a sufficient dataset, we plan to conduct assessments related to health conditions such as fall risk. This study, however, primarily aims to develop methods for extracting and analyzing radar signals and comparing them with data from other sensors.

\section{Conclusion}
 This study compared mmWave radar with Kinect and wearable sensors for STS analysis. Our experiments demonstrated that all three sensors can reliably count the number of STS movements completed within a fixed time, with high ICC values for duration indicating excellent agreement. Analysis of specific joint movements revealed that the sensors showed reasonable consistency in tracking trunk movements but less consistency in measuring smaller segments, such as knee movements.
To improve the accuracy of these measurements in future research, we propose employing a high-precision motion capture system and instructing participants to adopt a consistent sitting posture at the start of data collection.
Moving forward, we will further validate our findings in older adult populations at risk of falling. Additionally, we plan to compare sensor performance in other fall risk assessments, such as the TUG test, to further explore the capabilities of mmWave radar.
This study also highlights the advantages and limitations of individual sensors and suggests the potential of integrating sensor technologies to enhance the accuracy and reliability of human motion analysis.

\section*{Acknowledgment}
This work was supported by National Institute of Biomedical Imaging and Bioengineering (NIBIB), National Institute of Health (NIH), Grant No. R21EB033454.

\bibliographystyle{IEEEtran}
\bibliography{ref}

\begin{thebibliography}{10}
\providecommand{\url}[1]{#1}
\csname url@samestyle\endcsname
\providecommand{\newblock}{\relax}
\providecommand{\bibinfo}[2]{#2}
\providecommand{\BIBentrySTDinterwordspacing}{\spaceskip=0pt\relax}
\providecommand{\BIBentryALTinterwordstretchfactor}{4}
\providecommand{\BIBentryALTinterwordspacing}{\spaceskip=\fontdimen2\font plus
\BIBentryALTinterwordstretchfactor\fontdimen3\font minus \fontdimen4\font\relax}
\providecommand{\BIBforeignlanguage}[2]{{%
\expandafter\ifx\csname l@#1\endcsname\relax
\typeout{** WARNING: IEEEtran.bst: No hyphenation pattern has been}%
\typeout{** loaded for the language `#1'. Using the pattern for}%
\typeout{** the default language instead.}%
\else
\language=\csname l@#1\endcsname
\fi
#2}}
\providecommand{\BIBdecl}{\relax}
\BIBdecl

\bibitem{bohannon2010sit}
R.~W. Bohannon \emph{et~al.}, ``Sit-to-stand test: Performance and determinants across the age-span,'' \emph{Isokinetics and Exercise Science}, vol.~18, no.~4, pp. 235--240, 2010.

\bibitem{matthew2019estimating}
R.~P. Matthew \emph{et~al.}, ``Estimating sit-to-stand dynamics using a single depth camera,'' \emph{IEEE Journal of Biomedical and Health Informatics}, vol.~23, no.~6, pp. 2592--2602, 2019.

\bibitem{morgan2023automated}
C.~Morgan \emph{et~al.}, ``Automated real-world video analysis of sit-to-stand transitions predicts parkinson’s disease severity,'' \emph{Digital Biomarkers}, vol.~7, no.~1, pp. 92--103, 2023.

\bibitem{yan2024person}
K.~Yan \emph{et~al.}, ``Person-in-wifi 3d: End-to-end multi-person 3d pose estimation with wi-fi,'' in \emph{Proceedings of the IEEE/CVF Conference on Computer Vision and Pattern Recognition}, June 2024, pp. 969--978.

\bibitem{yang2022rethinking}
Z.~Yang \emph{et~al.}, ``Rethinking fall detection with wi-fi,'' \emph{IEEE Transactions on Mobile Computing}, vol.~22, no.~10, pp. 6126--6143, 2022.

\bibitem{jin2020mmfall}
F.~Jin \emph{et~al.}, ``mmfall: Fall detection using 4-d mmwave radar and a hybrid variational rnn autoencoder,'' \emph{IEEE Transactions on Automation Science and Engineering}, vol.~19, no.~2, pp. 1245--1257, 2020.

\bibitem{zhang2018real}
R.~Zhang and S.~Cao, ``Real-time human motion behavior detection via cnn using mmwave radar,'' \emph{IEEE Sensors Letters}, vol.~3, no.~2, pp. 1--4, 2019.

\bibitem{cuesta2010use}
A.~I. Cuesta-Vargas \emph{et~al.}, ``The use of inertial sensors system for human motion analysis,'' \emph{Physical Therapy Reviews}, vol.~15, no.~6, pp. 462--473, 2010.

\bibitem{poppe2007vision}
R.~Poppe, ``Vision-based human motion analysis: An overview,'' \emph{Computer Vision and Image Understanding}, vol. 108, no. 1-2, pp. 4--18, 2007.

\bibitem{wang2014quantitative}
F.~Wang \emph{et~al.}, ``Quantitative gait measurement with pulse-doppler radar for passive in-home gait assessment,'' \emph{IEEE Transactions on Biomedical Engineering}, vol.~61, no.~9, pp. 2434--2443, 2014.

\bibitem{seifert2019toward}
A.-K. Seifert \emph{et~al.}, ``Toward unobtrusive in-home gait analysis based on radar micro-doppler signatures,'' \emph{IEEE Transactions on Biomedical Engineering}, vol.~66, no.~9, pp. 2629--2640, 2019.

\bibitem{saho2020evaluation}
K.~Saho \emph{et~al.}, ``Evaluation of higher-level instrumental activities of daily living via micro-doppler radar sensing of sit-to-stand-to-sit movement,'' \emph{IEEE Journal of Translational Engineering in Health and Medicine}, vol.~8, pp. 1--11, 2020.

\bibitem{vaidya2017sit}
T.~Vaidya \emph{et~al.}, ``Sit-to-stand tests for copd: A literature review,'' \emph{Respiratory Medicine}, vol. 128, pp. 70--77, 2017.

\bibitem{bohannon20191}
R.~W. Bohannon and R.~Crouch, ``1-minute sit-to-stand test: Systematic review of procedures, performance, and clinimetric properties,'' \emph{Journal of Cardiopulmonary Rehabilitation and Prevention}, vol.~39, no.~1, pp. 2--8, 2019.

\bibitem{hu2024mmpose}
S.~Hu \emph{et~al.}, ``mmpose-fk: A forward kinematics approach to dynamic skeletal pose estimation using mmwave radars,'' \emph{IEEE Sensors Journal}, vol.~24, no.~5, pp. 6469--6481, 2024.

\bibitem{whitney2005clinical}
S.~L. Whitney \emph{et~al.}, ``Clinical measurement of sit-to-stand performance in people with balance disorders: Validity of data for the five-times-sit-to-stand test,'' \emph{Physical Therapy}, vol.~85, no.~10, pp. 1034--1045, 2005.

\bibitem{nuzik1986sit}
S.~Nuzik \emph{et~al.}, ``Sit-to-stand movement pattern: A kinematic study,'' \emph{Physical Therapy}, vol.~66, no.~11, pp. 1708--1713, 1986.

\bibitem{schenkman1990whole}
M.~Schenkman \emph{et~al.}, ``Whole-body movements during rising to standing from sitting,'' \emph{Physical Therapy}, vol.~70, no.~10, pp. 638--648, 1990.

\bibitem{van2013automated}
R.~C. Van~Lummel \emph{et~al.}, ``Automated approach for quantifying the repeated sit-to-stand using one body-fixed sensor in young and older adults,'' \emph{Gait \& Posture}, vol.~38, no.~1, pp. 153--156, 2013.

\bibitem{cheng1998sit}
P.-T. Cheng \emph{et~al.}, ``The sit-to-stand movement in stroke patients and its correlation with falling,'' \emph{Archives of Physical Medicine and Rehabilitation}, vol.~79, no.~9, pp. 1043--1046, 1998.

\bibitem{kera2020association}
T.~Kera \emph{et~al.}, ``Association between ground reaction force in sit-to-stand motion and falls in community-dwelling older japanese individuals,'' \emph{Archives of Gerontology and Geriatrics}, vol.~91, p. 104221, 2020.

\bibitem{mao2018crucial}
Y.~R. Mao \emph{et~al.}, ``The crucial changes of sit-to-stand phases in subacute stroke survivors identified by movement decomposition analysis,'' \emph{Frontiers in Neurology}, vol.~9, p. 185, 2018.

\bibitem{mentiplay2020five}
B.~F. Mentiplay \emph{et~al.}, ``Five times sit-to-stand following stroke: Relationship with strength and balance,'' \emph{Gait \& Posture}, vol.~78, pp. 35--39, 2020.

\bibitem{crook2017multicentre}
S.~Crook \emph{et~al.}, ``A multicentre validation of the 1-min sit-to-stand test in patients with copd,'' \emph{European Respiratory Journal}, vol.~49, no.~3, 2017.

\bibitem{medina2022prognostic}
F.~Medina-Mirapeix \emph{et~al.}, ``Prognostic value of the five-repetition sit-to-stand test for mortality in people with chronic obstructive pulmonary disease,'' \emph{Annals of Physical and Rehabilitation Medicine}, vol.~65, no.~5, p. 101598, 2022.

\bibitem{spence2023one}
J.~G. Spence \emph{et~al.}, ``One-minute sit-to-stand test as a quick functional test for people with copd in general practice,'' \emph{NPJ Primary Care Respiratory Medicine}, vol.~33, no.~1, p.~11, 2023.

\bibitem{boswell2023smartphone}
M.~A. Boswell \emph{et~al.}, ``Smartphone videos of the sit-to-stand test predict osteoarthritis and health outcomes in a nationwide study,'' \emph{npj Digital Medicine}, vol.~6, no.~1, p.~32, 2023.

\bibitem{yee2021performance}
X.~S. Yee \emph{et~al.}, ``Performance on sit-to-stand tests in relation to measures of functional fitness and sarcopenia diagnosis in community-dwelling older adults,'' \emph{European Review of Aging and Physical Activity}, vol.~18, pp. 1--11, 2021.

\bibitem{bowser2015sit}
B.~Bowser \emph{et~al.}, ``Sit-to-stand biomechanics of individuals with multiple sclerosis,'' \emph{Clinical Biomechanics}, vol.~30, no.~8, pp. 788--794, 2015.

\bibitem{zheng2023validity}
P.~Zheng \emph{et~al.}, ``Validity of the 30-second sit-to-stand test as a measure of lower extremity function in persons with multiple sclerosis: Preliminary evidence,'' \emph{Multiple Sclerosis and Related Disorders}, vol.~71, p. 104552, 2023.

\bibitem{bowman2023feasibility}
A.~Bowman \emph{et~al.}, ``Feasibility and safety of the 30-second sit-to-stand test delivered via telehealth: An observational study,'' \emph{PM\&R}, vol.~15, no.~1, pp. 31--40, 2023.

\bibitem{van2023self}
S.~J.~M. van Cappellen-van Maldegem \emph{et~al.}, ``Self-performed five times sit-to-stand test at home as (pre-) screening tool for frailty in cancer survivors: Reliability and agreement assessment,'' \emph{Journal of Clinical Nursing}, vol.~32, no. 7-8, pp. 1370--1380, 2023.

\bibitem{bergen2019cdc}
\BIBentryALTinterwordspacing
G.~Bergen and I.~Shakya, ``{CDC STEADI}: Evaluation guide for older adult clinical fall prevention programs,'' 2019. [Online]. Available: \url{https://www.cdc.gov/steadi}
\BIBentrySTDinterwordspacing

\bibitem{tulipani2022evaluation}
L.~J. Tulipani \emph{et~al.}, ``Evaluation of unsupervised 30-second chair stand test performance assessed by wearable sensors to predict fall status in multiple sclerosis,'' \emph{Gait \& Posture}, vol.~94, pp. 19--25, 2022.

\bibitem{wang2023high}
K.~Wang \emph{et~al.}, ``High accuracy machine learning model for sarcopenia severity diagnosis based on sit-to-stand motion measured by two micro motion sensors,'' \emph{medRxiv}, pp. 2023--05, 2023.

\bibitem{saho2021screening}
K.~Saho \emph{et~al.}, ``Screening of apathetic elderly adults using kinematic information in gait and sit-to-stand/stand-to-sit movements measured with doppler radar,'' \emph{Health Informatics Journal}, vol.~27, no.~1, p. 1460458221990051, 2021.

\bibitem{soubra2023automation}
R.~Soubra \emph{et~al.}, ``Automation of the timed up and go test using a doppler radar system for gait and balance analysis in elderly people,'' \emph{Journal of Healthcare Engineering}, vol. 2023, no.~1, p. 2016262, 2023.

\bibitem{yoon2024detection}
H.~Yoon and H.-C. Shin, ``Detection of fall risk behaviors in patients with severe mobility issues using fmcw radar: Sitting up and sitting on the side of the bed,'' \emph{Journal of Electromagnetic Engineering and Science}, vol.~24, no.~1, pp. 65--77, 2024.

\bibitem{hu2024radar}
S.~Hu \emph{et~al.}, ``Radar-based fall detection: A survey,'' \emph{IEEE Robotics \& Automation Magazine}, vol.~31, no.~3, pp. 170--185, 2024.

\bibitem{gurbuz2024overview}
S.~Z. Gurbuz \emph{et~al.}, ``Overview of radar-based gait parameter estimation techniques for fall risk assessment,'' \emph{IEEE Open Journal of Engineering in Medicine and Biology}, vol.~5, pp. 735--749, 2024.

\bibitem{sengupta2020mm}
A.~Sengupta \emph{et~al.}, ``mm-pose: Real-time human skeletal posture estimation using mmwave radars and cnns,'' \emph{IEEE Sensors Journal}, vol.~20, no.~17, pp. 10\,032--10\,044, 2020.

\bibitem{sengupta2022mmpose}
A.~Sengupta and S.~Cao, ``mmpose-nlp: A natural language processing approach to precise skeletal pose estimation using mmwave radars,'' \emph{IEEE Transactions on Neural Networks and Learning Systems}, vol.~34, no.~11, pp. 8418--8429, 2022.

\bibitem{hu2022stabilizing}
S.~Hu \emph{et~al.}, ``Stabilizing skeletal pose estimation using mmwave radar via dynamic model and filtering,'' in \emph{2022 IEEE-EMBS International Conference on Biomedical and Health Informatics}.\hskip 1em plus 0.5em minus 0.4em\relax IEEE, 2022, pp. 1--6.

\bibitem{toosizadeh2020effect}
N.~Toosizadeh \emph{et~al.}, ``The effect of vibratory stimulation on the timed-up-and-go mobility test: A pilot study for sensory-related fall risk assessment,'' \emph{Physiological Research}, vol.~69, no.~4, pp. 721--726, 2020.

\end{thebibliography}

\end{document}